\documentclass[aps,prx,twocolumn,superscriptaddress,longbibliography]{revtex4-2}
\usepackage{graphicx}
\usepackage{dcolumn}
\usepackage{bm}
\usepackage{bibunits}
\usepackage{upgreek}

\usepackage{physics}
\usepackage{svg}
\usepackage{placeins}

\newcommand{\eqlab}[1]{\label{eq:#1}}
\renewcommand{\eqref}[1]{Eq.~(\ref{eq:#1})}
\newcommand{\eqsref}[2]{Eqs.~(\ref{eq:#1}) and~(\ref{eq:#2})}

\newcommand{\figref}[1]{Fig.~\ref{fig:#1}}

\newcommand{\figlab}[1]{\label{fig:#1}}

\newcommand{\secref}[1]{Section~\ref{sec:#1}}
\newcommand{\secsref}[2]{Sections~\ref{sec:#1} and~\ref{sec:#2}}
\newcommand{\seclab}[1]{\label{sec:#1}}

\usepackage[
  colorlinks=true,
  linkcolor=blue,
  citecolor=blue,
  urlcolor=blue
]{hyperref}

\begin{document}

\preprint{APS/123-QED}

\title{Recirculating Quantum Photonic Networks for Fast Deterministic Quantum Information Processing}

\author{Emil Grovn}%
\thanks{Contact author: e.grovn@gmail.com}
\affiliation{DTU Electro, Technical University of Denmark, Building 343, 2800 Kgs. Lyngby, Denmark}%
\affiliation{NanoPhoton - Center for Nanophotonics, Technical University of Denmark, Building 343, 2800 Kgs. Lyngby, Denmark}

\author{Matias Bundgaard-Nielsen}%
\affiliation{DTU Electro, Technical University of Denmark, Building 343, 2800 Kgs. Lyngby, Denmark}%
\affiliation{NanoPhoton - Center for Nanophotonics, Technical University of Denmark, Building 343, 2800 Kgs. Lyngby, Denmark}

\author{Jesper M\o rk}%
\affiliation{DTU Electro, Technical University of Denmark, Building 343, 2800 Kgs. Lyngby, Denmark}%
\affiliation{NanoPhoton - Center for Nanophotonics, Technical University of Denmark, Building 343, 2800 Kgs. Lyngby, Denmark}

\author{Dirk Englund}
\affiliation{Department of Electrical Engineering and Computer Science, Massachusetts Institute of Technology,
77 Massachusetts Avenue, Cambridge, Massachusetts 02139, USA}

\author{Mikkel Heuck}
\thanks{Contact author: mheu@dtu.dk}
\affiliation{DTU Electro, Technical University of Denmark, Building 343, 2800 Kgs. Lyngby, Denmark}
\affiliation{NanoPhoton - Center for Nanophotonics, Technical University of Denmark, Building 343, 2800 Kgs. Lyngby, Denmark}

\date{\today}

\begin{abstract}
A fundamental challenge in deterministic photonic quantum information processing (QIP) is to realize key transformations on time scales shorter than those of detrimental decoherence and loss mechanisms.
Much work has addressed this challenge through device-focused approaches that aim to increase nonlinear interactions relative to loss rates. 
In this work, we adopt a complementary architecture-focused approach that minimizes process duration by directly realizing target transformations through hardware-native multi-photon dynamics, rather than decomposing them into standard gates.
We introduce a recirculating quantum photonic network (RQPN) in which photons are recirculated between nonlinear cavities with dynamically controlled waveguide couplings.
We demonstrate the RQPN's architectural advantage through two examples. First, we show that processing all qubits simultaneously yields a shorter duration than single- and two-qubit decompositions of the three-qubit Toffoli gate. Second, we demonstrate implementations of measurement-free single-photon loss correction, achieving up to seven-fold speedups and substantially reduced component counts relative to state-of-the-art photonic architecture proposals. Our results show that direct realization of QIP tasks, exploiting hardware-native multi-photon dynamics, can enable substantial reductions in process duration. This relaxes the required nonlinear interaction rates, thereby lowering the barrier to experimental realization of deterministic photonic QIP. 
\end{abstract}

\maketitle

\section{Introduction}
Photons can travel over long distances with low loss and decoherence in optical fibers, making photonic qubits an excellent candidate for, e.g., secure communication and distributed quantum computing~\cite{kimble2008quantum}.
To perform universal qubit operations, both single- and two-qubit gates are needed~\cite{nielsen2010quantum}. With dual-rail encoding, single-qubit gates are easily realized using linear optical elements~\cite{kok2007linear}, while two-qubit entangling gates require nonlinear transformations of photonic modes. Measurement-induced effective nonlinearities are being pursued for universal quantum computing~\cite{knill2001scheme,kok2007linear,bartolucci2023fusion}.
However, the resulting probabilistic gates require multiplexing, leading to a significant resource overhead. 

Deterministic entangling gates based on matter-mediated optical nonlinearities are challenged by fundamental limitations related to wave packet distortions~\cite{shapiro2006single,gea2010impossibility,xu2013analytic,nysteen2017limitations} and small nonlinear interaction rates in materials used for photonic integrated circuits~\cite{kok2007linear,slussarenko2019photonic}. These theoretical challenges have been addressed through device-focused approaches to implementing high-fidelity controlled-sign ($CZ$) gates~\cite{xia2016cavity,viswanathan2018analytical,levy2025passive,brod2016passive,ralph2015photon,yang2022deterministic,heuck2020pra,heuck2020prl,krastanov2022dyncav}. Additionally, impressive experimental demonstrations have improved the strength of nonlinear interaction rates for atom-like systems relative to loss and decoherence rates~\cite{Thompson1992, Hennessy2007, lodahl2017quantum, ota2018a}, although not yet to levels that enable high-fidelity $CZ$ gates~\cite{krastanov2022dyncav}.

The challenges of deterministic two-qubit logic exacerbate the difficulties of realizing complex quantum information processing (QIP) transformations that require circuits with many gates. 
This motivates a complementary architecture-focused route: rather than using a physical decomposition into single- and two-qubit gates, one may seek to realize the full target transformation directly through the native many-photon dynamics of the hardware.

Several recent photonic architectures have already taken this route, including architectures based on neural networks~\cite{steinbrecher2019quantum,ewaniuk2023imperfect,basani2025universal} and compact multimode nonlinear processors~\cite{krastanov2021room}. However, a clear reduction in process duration, relative to conventional gate decompositions and relative to existing photonic architectures based on direct realization, has remained elusive. 

In this work, we show that direct realization can enable significant reductions in process duration using a recirculating quantum photonic network (RQPN) architecture. Using the three-qubit Toffoli gate as an example, we find a direct realization with a clear reduction in process duration relative to single- and two-qubit (or -qutrit) gate decompositions. Using measurement-free correction of single-photon loss as a second example, 
we show that the RQPN achieves reductions in process duration by factors of $7.1$ and $4.5$ relative to the state-of-the-art architectures of Refs.~\cite{steinbrecher2019quantum} and \cite{krastanov2021room}, respectively.

\begin{figure}[!t]
     \centering
         \includegraphics[width=0.99\columnwidth]{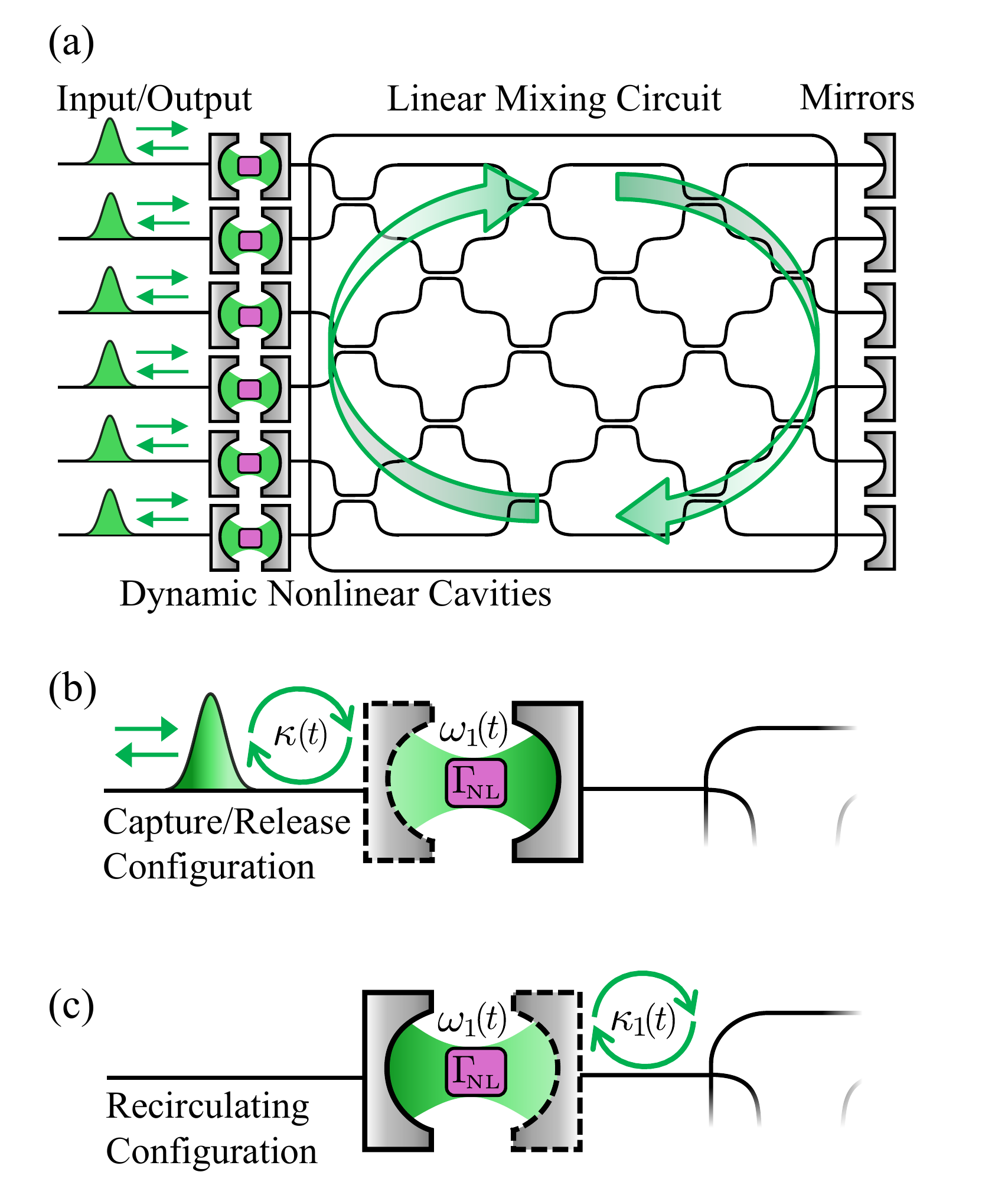}
\caption{(a) Sketch of the RQPN architecture. Dynamic nonlinear cavities are coupled to a linear mixing circuit composed of directional couplers, and the network is closed by mirrors on the right. The waveguide coupling of the nonlinear cavities is controllable. (b) To capture and release photonic quantum states from the cavities, they may be opened towards the input/output channel with a controllable coupling, $\kappa(t)$ (dashed outline of the mirror). (c) When a quantum state is trapped in the cavities, the RQPN is in its recirculating configuration, with the cavities closed towards the input/output channels (solid outline of the mirrors). In this configuration, photons interact in the cavities with a nonlinear interaction rate, $\Gamma_{\rm{NL}}$, and the controllable resonances, $\omega_m(t)$, and waveguide couplings, $\kappa_m(t)$, are optimized to implement a targeted QIP task. Note that (b) and (c) illustrate the top cavity ($m=1$). Practically, we envision the nonlinear cavities having only one mirror with a controllable coupling, and the configurations in (b) and (c) to be realized using fast routers~\cite{Heuck2023}.}  
\label{fig:RQPN}
\end{figure}

The RQPN operates by recirculating photons between its nonlinear cavities, steered by the dynamic control of the cavities.
The RQPN, illustrated in Fig.~\ref{fig:RQPN}(a), consists of dynamic nonlinear cavities connected to a linear mixing circuit composed of directional couplers. The nonlinear cavities have controllable resonance frequencies, $\omega_m(t)$, and waveguide couplings, $\kappa_m(t)$. This dynamic control enables input (output) quantum states to be captured (released) by the nonlinear cavities, as shown in Fig.~\ref{fig:RQPN}(b). After photon capture, the RQPN is in its recirculating configuration, where the dynamic control is used to steer the quantum state toward a targeted output by recirculating photons between the nonlinear cavities via the mixing circuit, as illustrated in Figs.~\ref{fig:RQPN}(a,c). 
Experimental demonstrations of such controllable cavities are found, e.g., in Refs.~\cite{tanaka2007dynamic, Xu2007, Nakadai2022, Zhang2019}, and details on their use for implementing deterministic $CZ$ gates in Refs.~\cite{heuck2020pra, heuck2020prl, krastanov2022dyncav}. Whereas previous architectures have used dynamic cavity coupling to capture and release traveling pulses to transform a photonic state via closed single-cavity evolution  \cite{heuck2020pra,heuck2020prl,krastanov2022dyncav,krastanov2021room}, the RQPN uses dynamic coupling to steer the evolution of a multi-cavity state.  Thus, we focus on optimizing the controls, $\omega_m(t)$ and $\kappa_m(t)$, to minimize the duration of QIP tasks with the RQPN in its recirculating configuration. In practical implementations, we envision the nonlinear cavities being connected to the input/output channels and mixing circuit via fast routers~\cite{Heuck2023} to switch between the capture/release and recirculating configurations. \\

This manuscript is organized as follows: Section~\ref{sec_model} presents the RQPN model and numerical framework used to optimize the controls. In Sec.~\ref{sec_tof}, we consider implementations of the three-qubit Toffoli gate based on self-phase modulation (SPM). Section~\ref{sec_MFCSP} presents our results on measurement-free implementations of single-photon loss correction using a bosonic encoding~\cite{chuang1997bosonic, Michael2016} with nonlinearities originating from SPM or interactions with two-level emitters (TLEs). Section~\ref{sec:Towards Experimental Realizations} contains a discussion that relates our results to experimentally demonstrated parameters from integrated photonics and optical nonlinearities. Finally, we conclude and provide an outlook for future work in~\secref{Conclusion and Outlook}.

\section{Model and Optimal Control}\label{sec_model}
We assume a perfect fidelity of the capture (release) of the input (output) quantum state of the RQPN before (after) the evolution in the recirculating configuration. This assumption is valid for sufficiently short input (output) pulses compared to the characteristic time scale of the nonlinear interactions, $1/\Gamma_{\rm NL}$ \cite{heuck2020pra,heuck2020prl,krastanov2022dyncav}. In the following, we thus neglect the duration of the capture and release processes and solely focus on describing and optimizing the dynamics of the RQPN in its recirculating configuration.
To describe the RQPN in~\figref{RQPN}, we use the SLH framework~\cite{gough2009, gough2009series, combes2017a} to only consider the degrees of freedom in the localized quantum systems (cavities). In the SLH framework, the influence on the cavity field from the continuum field of the waveguide is described as interactions with a broadband bosonic bath \cite{combes2017a,gardiner1985a,gardiner2004book}. We consider a frequency interval of waveguide modes centered around $\omega_0$ and use $\Omega_{\rm{int}}$ to denote the frequency window around $\omega_0$ over which the cavity-waveguide interaction can be approximated as frequency-independent, described by a Markovian coupling rate $\kappa_m$.
The Markov approximation of the SLH framework requires that this flat interaction window, $\Omega_{\rm{int}}$, is large compared to the characteristic rates of the localized system: $\kappa_m(t)$, $|\delta_m(t)|\!=\!|\omega_m(t)-\omega_0|$, and $\Gamma_{\rm NL}$, as well as the modulation bandwidth of the controls $\Omega_{\rm mod}$, which is defined in Eq.~(S22) of the Supplemental Material~\cite{grovn2026supplementary}.  
Finally, the propagation time of the waveguide fields across the mixing circuit, $T_{\rm{prop}}$, is assumed to be negligible in the SLH framework, which is valid for $1/T_{\rm{prop}}\gg \kappa_m(t),\,\Gamma_{\rm NL},$ and $\Omega_{\rm mod}$. In~\secref{Towards Experimental Realizations}, we relate our model assumptions to the experimental parameters of state-of-the-art photonic integrated circuits.

\subsection{Recirculating Circuit Model}\seclab{Recirculating Circuit Model}
In its recirculating configuration, the RQPN is described by the Hamiltonian (Supplemental Material SI~\cite{grovn2026supplementary}):
\begin{multline}\eqlab{RQPN Hamiltonian}
    \hat{H}(t) = \hbar\sum_{m}\delta^c_m(t)\hat{a}_m ^{\dagger}\hat{a}_m \\
    + \hbar \sum_n\sum_m C_{nm}\sqrt{\kappa_n^*(t)\kappa_m(t)}\hat{a}_n^{\dagger}\hat{a}_m + \hat{H}_{{\rm{NL}}}.
\end{multline}
In each sum, the index runs from $1$ to $M$, where $M$ is the number of cavities.
Each cavity, $m$, is considered to have a single mode with annihilation operator $\hat{a}_m$ and creation operator $\hat{a}_m^\dagger$. The corresponding $n$-photon state is denoted $\ket{n_m}$. We use a rotating frame oscillating at $\omega_0$ such that $\delta_m^c(t)=\omega_m^c(t)-\omega_0$, where $\omega_m^c(t)$ is the controllable resonance of each cavity. We added the superscript $c$ to distinguish cavity resonances from TLE transition frequencies introduced below. 
The linear mixing circuit couples all the cavities with controllable rates, $C_{nm}\sqrt{\kappa_n^*(t)\kappa_m(t)}$, where $\kappa_m(t)$ is the cavity-waveguide coupling of cavity $m$, assumed to be real and positive. The coupling matrix, $\bm{C}$, is related to the scattering matrix, $\bm{S}$, that defines the unitary transformation of modes when going from left to right through the mixing circuit and can be constructed with a mesh of directional couplers~\cite{reck1994,Clements:16}. The relation between $\boldsymbol{C}$ and $\bm{S}$ is
\begin{equation}\label{eq:sec2:C} 
    \bm{C}=\Im{\bm{S}\bm{S}^{\rm{T}}\big(\bm{I}-\bm{S}\bm{S}^{\rm{T}}\big)^{-1}},  
\end{equation}
where $\Im$ denotes the (elementwise) imaginary part and $\bm{I}$ is the identity matrix.
Since $\bm{S}$ can be any unitary matrix, we prove in Supplemental Material SII~\cite{grovn2026supplementary} that $\bm{C}$ can be any real symmetric matrix. 
The nonlinear Hamiltonian, $\hat{H}_{\rm{NL}}$, describes the optically nonlinear internal dynamics of the cavities. We consider, separately, the two cases of SPM
\begin{align}\eqlab{SPM Hamiltonian}
       \hat{H}_{\rm SPM}&=-\hbar \chi_3\sum_m \hat{a}_m^{\dagger}\hat{a}_m^{\dagger}\hat{a}_m \hat{a}_m ,
\end{align}
and interactions with TLEs
\begin{align}\eqlab{JC Hamiltonian}
       \hat{H}_{\rm JC} &= \hbar g\sum_m\left(\hat{\sigma}_m \hat{a}_m^{\dagger}+\hat{\sigma}_m^{\dagger}\hat{a}_m\right).
\end{align}
The nonlinear interaction rate, $\Gamma_{\rm NL}$, is $\chi_3$ for SPM interactions and $g$ for the Jaynes-Cummings Hamiltonian~\cite{Jaynes1963}. For interactions with TLEs, we also consider the emitter transition energy to be controllable, as described by the Hamiltonian
\begin{align}\eqlab{TLE Hamiltonian}
       \hat{H}_{\rm TLE} &=\sum_m\hbar\delta_m^e(t)\hat{\sigma}_m^{\dagger}\hat{\sigma}_m,
\end{align}
where $\delta_m^e(t)=\omega_m^e(t)-\omega_0$, $\hat{\sigma}_m=\ket{g_m}\!\bra{e_m}$, and $\hat{\sigma}^\dagger_m=\ket{e_m}\!\bra{g_m}$, with $\ket{g_m}$ and $\ket{e_m}$ being the ground and excited state of the TLE in cavity $m$ and $\omega_m^e(t)$ being the controllable transition frequency. 

\subsection{Numerical Control Framework}\seclab{Numerical Control Framework}
In the RQPN simulation, we neglect loss and decoherence effects from the surrounding environment. We optimize the controls to minimize the quantum process duration, $T$, which, for entangling processes, is ultimately limited by the nonlinear coupling rate, $\Gamma_{\rm{NL}}$. To enable comparisons of time efficiency between different material platforms and architectures, we define a dimensionless process duration, $T_{\eta}$, by measuring the duration in units of the inverse nonlinear interaction rate, $T=T_{\eta}/\Gamma_{\rm{NL}}$. Minimizing $T_{\eta}$ lowers the requirement on the size of the nonlinear rate relative to the dominant decoherence rate, thereby lowering the barrier for experimental demonstrations of quantum information processing.

We specify a QIP task by a transformation of a set of input states to a targeted set of output states
\begin{align}\eqlab{Target Transformation}
    \ket{\psi^{(n)}(T)} = \hat{U}_{\rm{tar}}\ket{\psi_{\rm in}^{(n)}}, \quad n=1,...,N_{\rm task}.
\end{align}
The controls are time-signals parametrized as piecewise-constant functions over $N_{\rm{bin}}$ time bins~\cite{2022propson, Goerz2022quantumoptimal, Goldschmidt2022modelpredictive}. Thus, the mapping from the initial to the final state is
\begin{equation}\eqlab{RQPN transformation}
    \ket{\psi(T)}=\!\left(\hat{U}_{N_{\rm{bin}}}\ldots\hat{U}_2 \hat{U}_1\right)\ket{\psi_{\rm in}} = \hat{U}_{\rm{RQPN}} \ket{\psi_{\rm in}}\!,
\end{equation}
where
\begin{align}\eqlab{Up}
    \hat{U}_p=\exp\!\left(-i\hat{H}_p\Delta t_p/\hbar\right),
\end{align}
and $\hat{H}_p$ is the Hamiltonian of the $p$'th time bin.
We optimize the coupling rates $\kappa_m(t_p)$, detunings $\delta_m^c(t_p)$ and $\delta_m^e(t_p)$ ($t_p$ being the time in the $p$'th bin), time bin durations $\Delta t_p$ , and the coupling matrix $\bm{C}$ (see Supplemental Material SIII~\cite{grovn2026supplementary} for parameterization details). 

The time evolution in~\eqsref{RQPN transformation}{Up} is evaluated numerically in Python using the open-source framework \texttt{Dynamiqs}~\cite{guilmin2025dynamiqs}, which utilizes \texttt{Jax} for the automatic differentiation back-end~\cite{jax2018github}. We use the gradient-descent-based algorithm \texttt{Adam}~\cite{Goodfellow-et-al-2016} to minimize a cost function, $\mathcal{E}(\bm{\theta})$, which depends on the vector $\bm{\theta}$ containing all the trainable parameters (see Supplemental Material SIII~\cite{grovn2026supplementary} for details of the optimization procedure). The primary optimization objective is to maximize the overlap between the output and target states for all $N_{\rm{task}}$ inputs specified by the QIP task under consideration. Therefore, the cost function contains a term $\mathcal{E}_{\mathcal{I}} = \ln(\mathcal{I})$, with the infidelity $\mathcal{I}$ defined by
\begin{align}\eqlab{cost infidelity}
    \mathcal{I}=1- \left|\frac{1}{N_{\rm task}}\sum_{n=1}^{N_{\rm task}}\bra{\psi_{\rm in}^{(n)}}\hat{U}_{\rm{tar}}^\dagger \hat{U}_{\rm{RQPN}}\ket{\psi_{\rm in}^{(n)}}\right|^2 \!\!.
\end{align}
We stop the optimization once $\mathcal{I}$ drops below a threshold, $\mathcal{I}_{\rm{th}}$, typically set to $\mathcal{I}_{\rm{th}}=0.1\,\%$. Additional terms may be added to $\mathcal{E}$ to penalize large control bandwidths, while the control amplitudes are explicitly bounded by our choice of parameterization, as described in Supplemental Material SIII~\cite{grovn2026supplementary}. The time steps $\Delta t_p$, and thereby the duration $T=\sum_{p=1}^{N_{\rm bin}}\Delta t_p$, are also explicitly bounded from below and above. To minimize the process duration, we consider many different optimization configurations and gradually reduce the upper bound for each configuration until $\mathcal{I}\leq\mathcal{I}_{\rm{th}}$ cannot be achieved.

\section{Advantage of Multi-Qubit Processing}\label{sec_tof}
The coupling matrix, $\bm{C}$, can be chosen so that the RQPN exhibits all-to-all coupling between the nonlinear cavities throughout the time evolution of the recirculating configuration. Therefore, for a given QIP task, we can search for a direct implementation of the full transformation rather than decomposing it into smaller transformations, each involving fewer photons and modes. To illustrate the advantage of this all-to-all connectivity and multi-qubit processing, we consider the three-qubit Toffoli gate on dual-rail encoded qubits, implemented using the SPM interactions described by the Hamiltonian in~\eqref{SPM Hamiltonian}. We compare a direct implementation to decompositions into single- and two-qubit (or qutrit) gates.

\subsection{Decomposition of the Toffoli Gate using Single- and Two-Qubit Gates}
Single-qubit gates along with the two-qubit $CZ$ gate provide a universal gate set, which means that any $N$-qubit unitary operation can be decomposed using these gates~\cite{nielsen2010quantum}.
In dual-rail encoding, any single-qubit gate can be realized by linear transformations of optical modes and is thus controlled by the cavity-cavity coupling term and cavity detuning term in \eqref{RQPN Hamiltonian}. Their duration, i.e., the time required to transform the input state into the desired output state, is limited only by the amplitude and bandwidth of $\kappa_m(t)$ and $\delta_m^c(t)$, and is negligible in practice where the SPM strength, $\chi_3$, is the limiting physical property of the system.
On the other hand, two-qubit entangling gates, such as the $CZ$ gate, require nonlinear transformations of optical modes~\cite{kok2007linear}, so their duration is limited by $\chi_3$. In Supplemental Material SIV A~\cite{grovn2026supplementary}, we prove that, when using SPM nonlinearities, the $CZ$-gate duration is bounded from below by $(\pi/4)/\chi_3$ for an infidelity of $\mathcal{I}=0$ and by $0.754/\chi_3$ for $\mathcal{I} = 0.1\%$. We find both an analytical implementation with this duration and a numerical RQPN implementation with a duration of $0.773/\chi_3$ for $\mathcal{I} = 0.1\%$. The minimum number of $CZ$ gates required to implement the three-qubit Toffoli gate is six~\cite{Ralph2007}. 
However, if an additional optical mode is available, one of the qubits may be represented as a qutrit with three logical states $\{\ket{0}_{\!L}\!,~\ket{1}_{\!L}\!,~\ket{2}_{\!L} \}$. A $CZ$ gate operating on a qubit and a qutrit, defined by an operator $\hat{U}_{CZ}$, causes the following transformations \cite{Ralph2007}
\begin{equation}\eqlab{CZ qubit-qutrit}
\begin{split}
    \hat{U}_{CZ}\ket{0,0}_{\!L}=\ket{0,0}_{\!L},&\quad \hat{U}_{CZ}\ket{0,1}_{\!L}=\ket{0,1}_{\!L},\\ \hat{U}_{CZ}\ket{1,0}_{\!L}=\ket{1,0}_{\!L},&\quad \hat{U}_{CZ}\ket{1,1}_{\!L}=-\ket{1,1}_{\!L}, \\
    \hat{U}_{CZ}\ket{0,2}_{\!L}=\ket{0,2}_{\!L},&\quad \hat{U}_{CZ}\ket{1,2}_{\!L}=\ket{1,2}_{\!L} .
\end{split}
\end{equation}
With access to an additional mode and the qubit-qutrit $CZ$ operator, $\hat{U}_{CZ}$, a decomposition of the Toffoli gate only requires three qubit-qutrit $CZ$ gates \cite{Ralph2007}. In Supplemental Material IV B~\cite{grovn2026supplementary}, we prove that the duration of $\hat{U}_{CZ}$ is bounded from below by $(\pi/4)/\chi_3$ for $\mathcal{I}=0$ and $0.754/\chi_3$ for $\mathcal{I}=0.1\%$, but the existence of an implementation with this duration is not guaranteed. Numerically, we find an RQPN implementation of the qubit-qutrit $CZ$ gate with a duration of $0.89/\chi_3$ for $\mathcal{I}=0.1\%$, which, to the best of our knowledge, is the fastest existing implementation with SPM interactions. The corresponding Toffoli gate duration is $2.67/\chi_3$. 
Operating on two photons in five modes rather than four modes provides a significant speedup of the three-qubit Toffoli gate, in practice, from $4.64/\chi_3$ (with $\mathcal{I}=0.86\%$ for the full Toffoli gate) to $2.67/\chi_3$ (with $\mathcal{I}=0.30\%$ for the full Toffoli gate) using qubit-qubit and qubit-qutrit $CZ$ gates with infidelity $\mathcal{I}=0.1\%$, respectively. In the next section, we investigate the potential speedup when operating on all three photons in all six modes simultaneously using the RQPN architecture. 

\begin{table}[]
    \centering
    \setlength{\extrarowheight}{3pt}
    \begin{tabular}{|c|c|c|}
        \hline 
        Qubit decom.  & Qutrit decom. & Direct \\
        \hline
        $4.64/\chi_3$ ($0.86\,\%$)  & $2.67/\chi_3$ ($0.30\,\%$) & $2.00/\chi_3$ ($<0.30\,\%$) \\
        \hline
     \end{tabular}
    \caption{Duration (infidelity) of numerical implementations of the Toffoli gate using a decomposition with qubit-qubit $CZ$ gates, qubit-qutrit $CZ$ gates, or direct realization.}
    \label{tab:toffoli_results}
\end{table}

\begin{figure}[h!]
     \centering
     \includegraphics[width=0.48\textwidth]{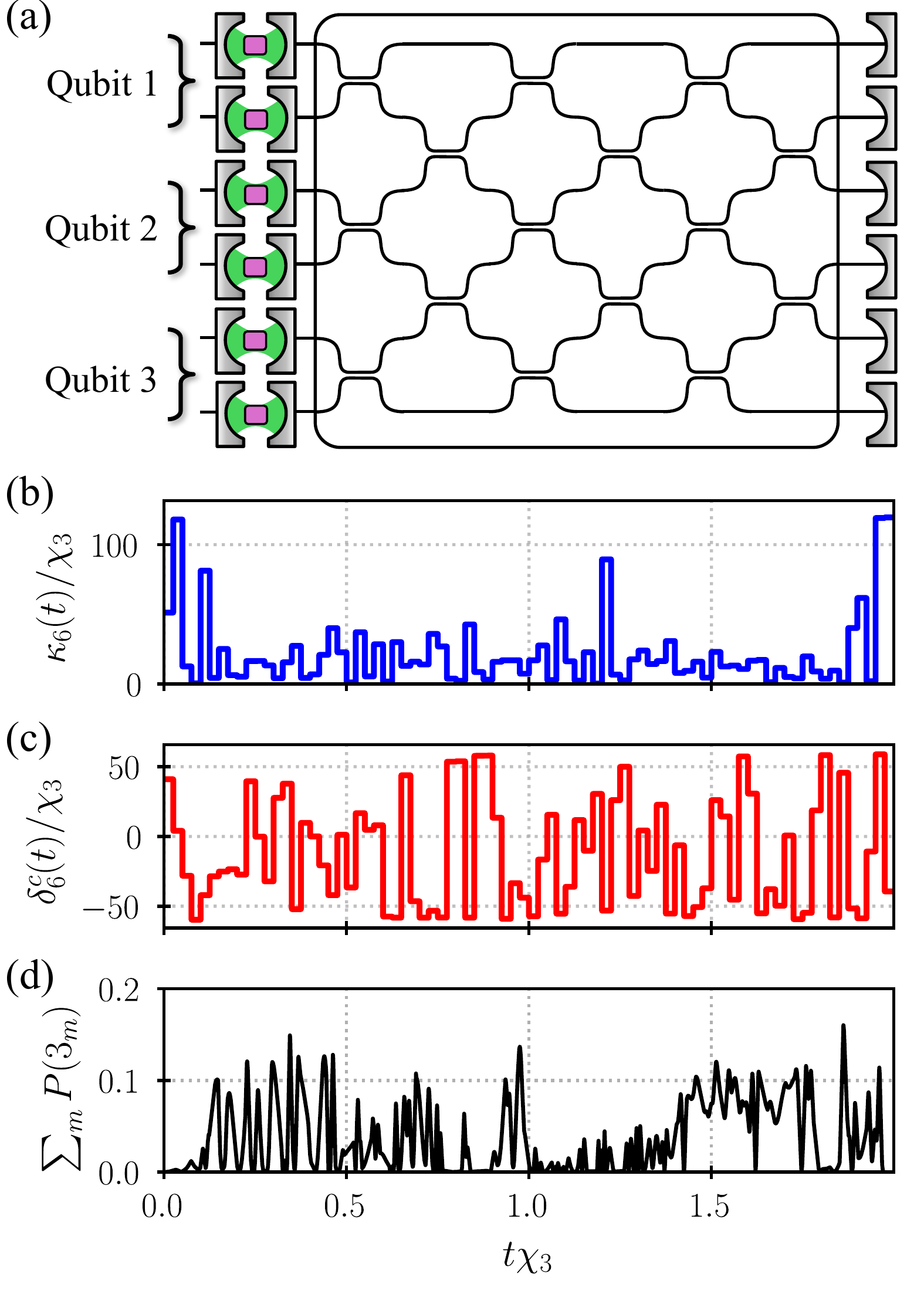}
\caption{Direct implementation of the three-qubit Toffoli gate with dual-rail encoding. (a) RQPN architecture with SPM nonlinearities and six cavities to encode three qubits. (b) Optimized cavity-waveguide coupling and (c) detuning for cavity $m=6$. (d) Probability of three photons occupying the same cavity, for the initial state $\ket{\psi_{\rm in}} = \ket{0,0,0}_L$. The probability of $n$ photons occupying cavity $m$ is $P(n_m) = \abs{\bra{\psi(t)}\ket{n_m}}^2$. 
The result was achieved using $N_{\rm{bin}}=80$, a fixed duration $T=2.0/\chi_3$, and trainable cavity-waveguide coupling, detuning, and coupling matrix $\bm{C}$. Additional information for the optimization and result is in Supplemental Material SV~\cite{grovn2026supplementary}, Fig.~6.}
\figlab{Toffoli}
\end{figure}

\subsection{Direct Implementation of the Toffoli Gate}
The RQPN used to implement the Toffoli gate is illustrated in~\figref{Toffoli}(a). We optimize the coupling matrix, $\bm{C}$, and the controls in the recirculating configuration with $N_{\rm bin}=80$ to implement the Toffoli gate.
Figure~\ref{fig:Toffoli}(b) and (c) show the control fields for cavity $m=6$, as an example, from an optimization resulting in $\mathcal{I}<0.30\%$ - see Supplemental Material SV~\cite{grovn2026supplementary} for the controls of the remaining cavities, the optimized coupling matrix, and the state evolution for the eight logical basis states. The duration of the Toffoli gate is reduced to $T=2.00/\chi_3$ when using the RQPN to simultaneously process all three qubits using all six modes, beating the implementations with qubit-qubit $CZ$ gates and qubit-qutrit $CZ$ gates by a factor of $2.32$ and $1.34$, respectively - these results are compared in Table \ref{tab:toffoli_results}. We ran the optimization using a fixed duration of $T=2.00/\chi_3$ and without constraining the control amplitudes and bandwidths to beat the analytical lower bound of $2.32/\chi_3$ with qubit-qutrit $CZ$ gates. This demonstrates the fundamental time-reducing advantage of processing all photons in all modes simultaneously. Our optimization procedure does not exhaust the space of possible implementations, so it is possible that a faster RQPN implementation exists. 

For SPM interactions, $n$-photon states acquire a nonlinear phase with a rate $n(n-1)\chi_3$. Therefore, the maximum rate of the fully connected RQPN ($6\chi_3$) is three times larger than that of the decompositions ($2\chi_3$). Figure~\ref{fig:Toffoli}(d) shows the probability of three photons all occupying the same cavity for the logical input $\ket{0,0,0}_L$. We observe that the optimizer makes use of the stronger nonlinearity. 

\section{Measurement-Free Correction of Single-Photon Loss}\label{sec_MFCSP}
A key requirement for utility-scale quantum computing is the ability to correct errors \cite{campbell_roads_2017,2008Aharonov,google2025quantum}. Photon loss is the dominant error channel of most photonics-based systems \cite{slussarenko2019photonic} and a major obstacle for scalable photonic quantum computing architectures \cite{bartolucci2023fusion,chan2025practicalblueprintlowdepthphotonic}. Photon-loss correction is also crucial in quantum communication \cite{2014Muralidharan}. In the following, we correct for single-photon loss using the RQPN.
We consider the following multi-photon bosonic encoding~\cite{chuang1997bosonic}:
\begin{align}\eqlab{Bosonic Encoding}
    \ket{0}_{\!L} &=\frac{1}{\sqrt{2}}\left(\ket{4_1,0_2}+\ket{0_1,4_2} \right), \quad  \ket{1}_{\!L}=\ket{2_1,2_2},
\end{align}
where each logical state contains four photons distributed across two modes. Miatto \textit{et al.}~\cite{miatto2018hamiltonians} proposed a theoretical scheme that uses the encoding in~\eqref{Bosonic Encoding} for measurement-free correction of single-photon loss, also denoted a measurement-free one-way quantum repeater.
We are aware of only two proposals for implementations of this task: the first uses a neural network architecture~\cite{steinbrecher2019quantum} with SPM interactions, and the second uses a circuit of triply-resonant cavities with interactions based on second-harmonic generation~\cite{krastanov2021room}. In the following, we consider RQPN implementations and compare their time and hardware efficiency with these alternative architectures for multi-photon processing.

\begin{figure*}[t]
     \centering
     \includegraphics[width=\textwidth]{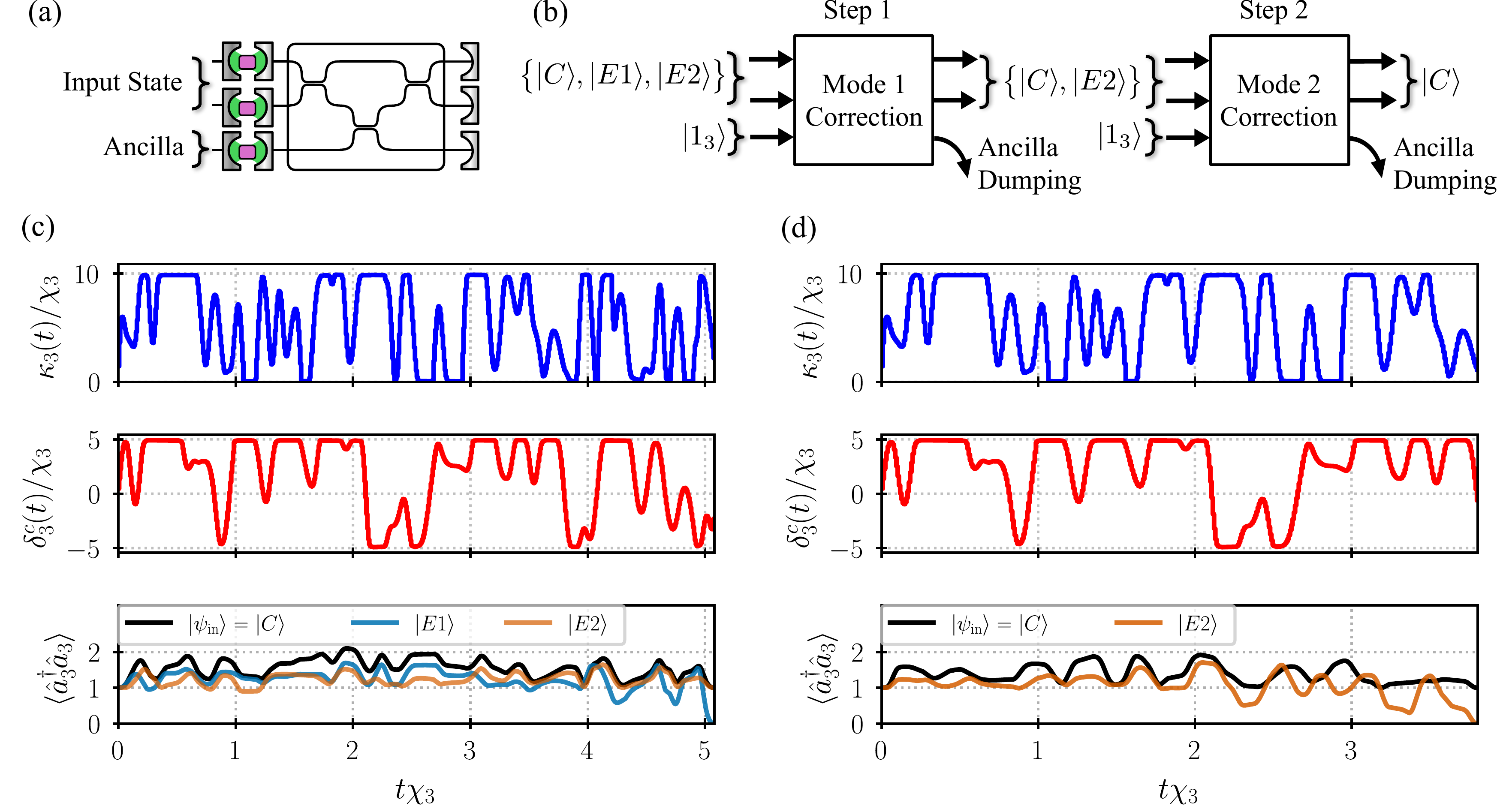}
\caption{Implementation of a measurement-free one-way quantum repeater using a three-mode RQPN with SPM interactions. (a) Modes 1 and 2 are used for the input state, and mode 3 contains the ancillary state. (b) The single-photon loss correction task is divided into two steps: the first corrects for loss in mode 1, and the second for loss in mode 2. Before each step, the ancilla cavity receives a single photon and is initialized to state $\ket{1_3}$. 
(c) Selected plots of the dynamics in step 1. The top and center panels plot the optimized waveguide coupling and detuning of the ancilla cavity ($m=3$). The bottom panel plots the expectation value of the photon number operator for the three correctable input states for the example of $\abs{\alpha}^2=\abs{\beta}^2=0.5$.
(d) Selected plots of the dynamics in step 2. In the bottom panel, only the input states $\ket{C}$ and $\ket{E2}$ are relevant. Additional information for the optimization and result is in Supplemental Material SVI~\cite{grovn2026supplementary}, Figs.~7 and 8.
}\figlab{Repeater SPM}
\end{figure*}

\subsection{Definition of the One-Way Repeater Task}
In measurement-free correction of single-photon loss, we consider a code state $\ket{C}=\alpha\ket{0}_{\!L}+\beta\ket{1}_{\!L}$ traveling through a lossy quantum channel. After propagation, the correctable error states are $\ket{E1}=\alpha\ket{3_1,0_2}+\beta\ket{1_1,2_2}$ and $\ket{E2}=\alpha\ket{0_1,3_2}+\beta\ket{2_1,1_2}$, where a photon was lost from mode 1 or 2, respectively. For a channel with, e.g., a $10\%$ risk of losing a single photon in either mode, there is a $1\%$ chance of uncorrectable multi-photon loss \cite{krastanov2021room}.
A unitary transformation maps orthogonal inputs to orthogonal outputs, so a measurement-free operation that maps any input $\ket{\psi_{\rm{in}}}=\{\ket{C}$, $\ket{E1}$, $\ket{E2}\}$ to the same output $\ket{C}$ requires an ancillary quantum system. The targeted transformation of the repeater is
\begin{align}\eqlab{repeater}
\begin{split}
\mathrm{No\,loss:}\quad&\ket{C}\,\,\,\otimes \ket{A0}\to \ket{C}\otimes \ket{A0} \\
    \mathrm{Mode\,1\,loss:}\quad&\ket{E1}\otimes \ket{A0}\to \ket{C}\otimes \ket{A1} \\
\mathrm{Mode\,2\,loss:}\quad&\ket{E2}\otimes \ket{A0}\to \ket{C}\otimes \ket{A2},
\end{split}
\end{align}
where $\ket{A0}$, $\ket{A1}$, and $\ket{A2}$ are orthogonal states of the ancilla. In~\secref{Repeater SPM}, we consider an implementation with SPM interactions and the initial ancilla state being photons in additional cavities. In~\secref{Repeater TLE}, we consider TLE interactions with the ancilla state being the TLE states.

\subsection{One-Way Repeater with SPM Interactions}\seclab{Repeater SPM}
For our repeater implementation with SPM interactions to be as hardware-efficient as possible, we employ a three-mode RQPN with a two-step protocol, utilizing one ancillary cavity that is initialized in a single-photon Fock state, $\ket{1_3}$, at the beginning of each step, as illustrated in Figs.~\ref{fig:Repeater SPM}(a) and~\ref{fig:Repeater SPM}(b).  
The first step corrects for photon loss in mode 1 by the transformation
\begin{align}\eqlab{SPM repeater step 1}
\begin{split}
\mathrm{No\,loss:}\quad&\ket{C}\,\,\,\otimes \ket{1_3}\to \ket{C}\otimes \ket{1_3} \\
    \mathrm{Mode\,1\,loss:}\quad&\ket{E1}\otimes \ket{1_3}\to \ket{C}\otimes \ket{0_3} \\
\mathrm{Mode\,2\,loss:}\quad&\ket{E2}\otimes \ket{1_3}\to \ket{E2}\otimes \ket{1_3}.
\end{split}
\end{align}
Before the second step, cavity 3 is emptied and re-initialized in the state $\ket{1_3}$, as illustrated in~\figref{Repeater SPM}(b). Therefore, correcting for loss in mode 2 in the second step imparts the transformation
\begin{align}\eqlab{SPM repeater step 2}
\begin{split}
\mathrm{No\,loss:}\quad&\ket{C}\,\,\,\otimes \ket{1_3}\to \ket{C}\otimes \ket{1_3} \\
\mathrm{Mode\,2\,loss:}\quad&\ket{E2}\otimes \ket{1_3}\to \ket{C}\otimes \ket{0_3}.
\end{split}
\end{align}
To simplify the optimization, we keep the coupling matrix fixed with values
\begin{equation}\eqlab{C matrix SPM repeater}
    \boldsymbol{C}=
    \begin{bmatrix}
0 & 1 & 1 \\
1 & 0 & 1 \\
1 & 1 & 0
\end{bmatrix}.
\end{equation}
Note that the diagonal elements can be set to zero, since the self-coupling terms, $C_{mm}\kappa_m(t)$, in~\eqref{RQPN Hamiltonian} may always be offset by shifting the detunings. Figures~\ref{fig:Repeater SPM}(c) and~\ref{fig:Repeater SPM}(d) plot the optimized controls and the expectation value of the photon number operator for the ancillary cavity, illustrating the donation of a photon for input state $\ket{E1}$ (c) and $\ket{E2}$ (d).
The corresponding process durations are $5.09/\chi_3$ for step 1 and $3.81/\chi_3$ for step 2. The duration of the full repeater task (still neglecting the duration of capture (release) before (after) each step) is $T=8.90/\chi_3$, which is $4.5$ and $7.1$ times faster than the reported implementations in Refs.~\cite{krastanov2021room} and~\cite{steinbrecher2019quantum}, respectively. We used $N_{\rm bin}=640$ and penalized the control bandwidths to obtain smooth signals at the expense of longer process duration (for instance, we found a solution with $T=8.36/\chi_3$ without bandwidth penalty and $N_{\rm bin}=80$). Each transformation was optimized separately to achieve $\mathcal{I}<0.05\%$. The resulting infidelity when simulating the full two-step process was $\mathcal{I}=0.08\%$.

\subsection{One-Way Repeater with TLE Interactions}\seclab{Repeater TLE}
To implement a repeater with TLE interactions, we design a compact two-mode RQPN by using the TLEs as ancillary systems, as illustrated in~\figref{Repeater TLE}(a).
\begin{figure}[t]
     \centering
          \includegraphics[width=\columnwidth]{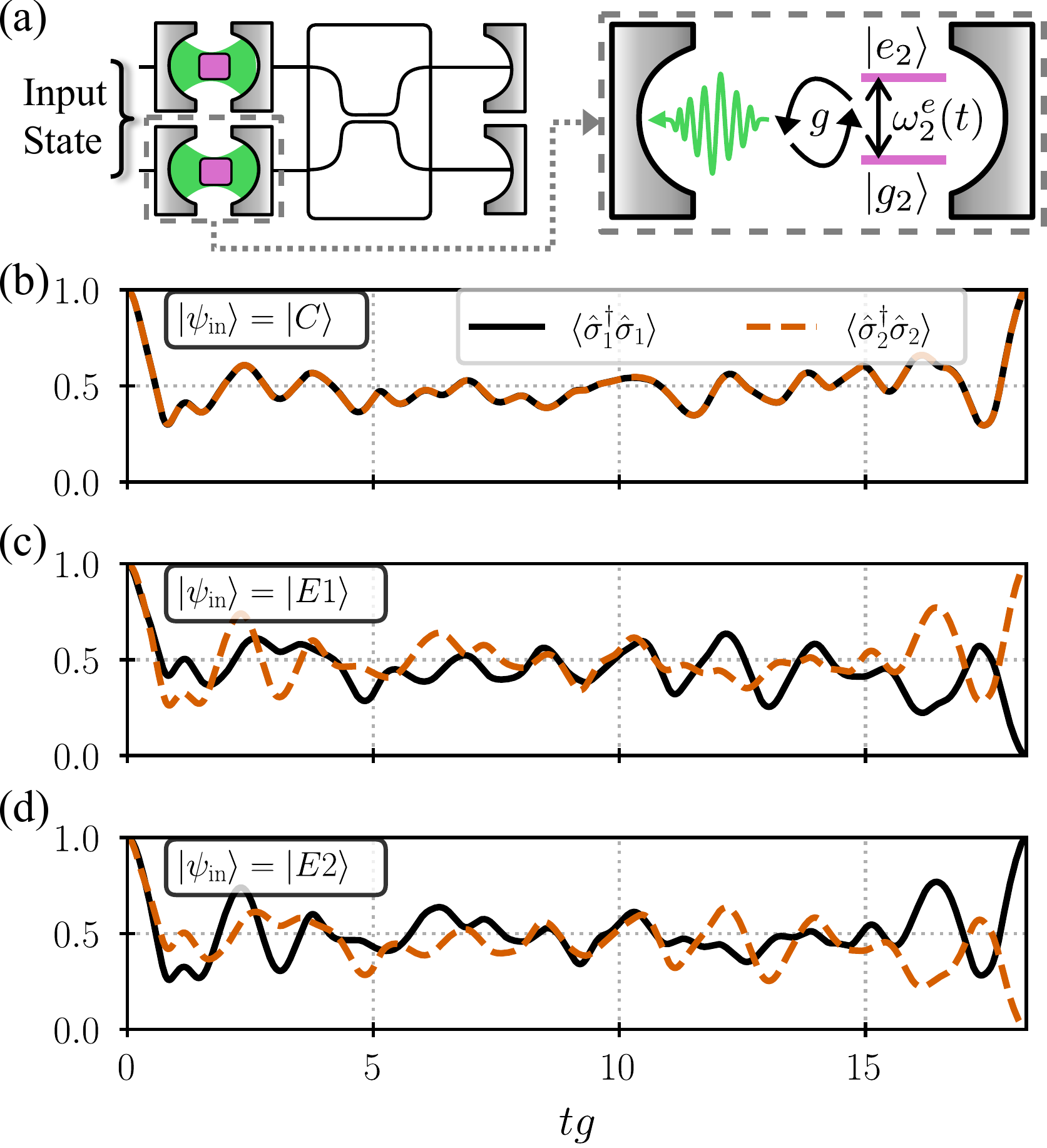}
\caption{Implementation of the measurement-free one-way quantum repeater using a two-mode RQPN with TLE interactions. (a) The input states are captured by the cavities, and the TLEs are used as ancillary systems by initializing them to their excited states, $\ket{A0}=\ket{e_1}\ket{e_2}$. (b)-(d) Plots of the evolution of the excited state probability of the two TLEs for input states $\ket{C}$ (b), $\ket{E1}$ (c), and $\ket{E2}$ (d). As in~\figref{Repeater SPM}, we consider the example of $\abs{\alpha}^2=\abs{\beta}^2=0.5$ in the input state. Additional information for the optimization and result, including optimized controls, is in Supplemental Material SVI~\cite{grovn2026supplementary}, Fig.~9.}
\figlab{Repeater TLE}
\end{figure}
The TLEs are initialized in the excited state, $\ket{e_m}$ for $m=1$ and $2$. The TLE in cavity $m$ donates a photon by relaxing to its ground state, if a photon is missing from mode $m$ in the photonic input state. Thus, the repeater transformation in~\eqref{repeater} is implemented with ancillary states $\ket{A0}=\ket{e_1}\!\ket{e_2}$, $\ket{A1}=\ket{g_1}\!\ket{e_2}$, and $\ket{A2}=\ket{e_1}\!\ket{g_2}$. We use a fixed coupling matrix
\begin{equation}
    \boldsymbol{C}=
    \begin{bmatrix}
0 & 1  \\
1 & 0 
\end{bmatrix},
\end{equation}
and optimize the cavity-waveguide couplings, cavity and TLE detunings, and time-bin durations. The resulting process duration is $T=18.31/g$, which is $2.2$ and $3.4$ times faster than the reported implementations in Refs.~\cite{krastanov2021room} and~\cite{steinbrecher2019quantum}. Figures~\ref{fig:Repeater TLE}(b-d) plot the excited state probability of TLE 1 (black line) and 2 (orange line) for input states $\ket{C}$ (b), $\ket{E1}$ (c), and $\ket{E2}$ (d). It is clearly observed how TLE $m$ donates a photon if a photon was lost from the corresponding input mode $m$. 
We used $N_{\rm bin}=640$ and penalized the control bandwidths to obtain smooth control signals and infidelity $\mathcal{I}<0.1\%$.

\subsection{Comparison of Architectures}\seclab{Comparison of Architectures}
Comparisons between different architectures require care, as they may rely on different material systems and pose different engineering challenges. Nevertheless, time and hardware efficiency may be compared using process durations in units of the inverse nonlinear interaction rate, $ T = T_\eta/ \Gamma_{\rm{NL}}$, and the required numbers of linear and nonlinear components. In Ref.~\cite{krastanov2021room}, second-order nonlinear interactions between two cavity modes, $\hat{a}$ and $\hat{b}$, were used with the Hamiltonian: $\hat{H}_{\rm{SHG}}=\hbar\chi_2(\hat{a}^\dagger\hat{b}\hat{b}+\hat{a}\hat{b}^\dagger\hat{b}^\dagger)$. The authors reported a one-way repeater duration of $40/\chi_2$ using 13 linear and 2 nonlinear components. In Ref.~\cite{steinbrecher2019quantum}, SPM interactions were used, described by the transformation $\sum_n^\infty e^{i\phi n(n-1)}\ket{n}\!\bra{n}$, with $\phi = \pi$. This transformation corresponds to a duration of $(\pi/2)/\chi_3$, when comparing to our SPM Hamiltonian in~\eqref{SPM Hamiltonian}. The neural network architecture in Ref.~\cite{steinbrecher2019quantum} required 40 layers, each layer consisting of a linear mixing circuit followed by nonlinear interactions, to implement the one-way repeater, resulting in a duration of $62.8/\chi_3$, 240 linear and 160 nonlinear components. Table~\ref{tab:repeater_results} summarizes these results along with the RQPN performance found in~\secsref{Repeater SPM}{Repeater TLE}.
\begin{table}[]
    \centering
    \setlength{\extrarowheight}{4pt}
    \begin{tabular}{|l|c|c|c|}
        \hline
        \multicolumn{4}{|c|}{\textbf{One-Way Repeater Implementations}}\\
        \hline 
        Architecture  & Duration & \# Linear  & \# Nonlinear  \\
        \hline
        RQPN (SPM) & $8.90/\chi_3$ & 3 & 3 \\
        \hline
        RQPN (TLE) &  $18.31/g$ & 1 & 2 \\
        \hline
       Neural Network \cite{steinbrecher2019quantum} &  $62.83/\chi_3$ & 240 & 160 \\
        \hline
        Three-mode cavities \cite{krastanov2021room} &  $40/\chi_2$ & 13 & 2\\
        \hline
     \end{tabular}
    \caption{Comparison of one-way repeater implementations. The columns show the duration, $T$, the number of linear components, and the number of nonlinear elements.}
    \label{tab:repeater_results}
\end{table}
The SPM-based RQPN and Ref.~\cite{steinbrecher2019quantum} use the same nonlinearity and thereby provide the clearest comparison. Table~\ref{tab:repeater_results} shows that the RQPN is $7.1$ times faster than the neural network architecture and reduces the component overhead by a factor of $80$ ($53.3$) for linear (nonlinear) elements. Compared to the three-mode processor architecture in Ref.~\cite{krastanov2021room}, the dimensionless duration, $T_\eta$, is $4.5$ times shorter for the RQPN with SPM interactions. Even with TLE interactions, where the dynamics are more complex, the RQPN outperforms existing architectures in time efficiency by factors of 2.2 and 3.4, as seen from Table~\ref{tab:repeater_results}, while using only one linear and two nonlinear components.

For each type of nonlinear interaction, we have found the ratio between the duration of RQPN implementations of the one-way repeater, $T_{\rm rep}$, and the dual-rail qubit-qubit $CZ$ gate, $T_{CZ}$. We obtain $T_{\rm rep}/T_{CZ}=11.5$ and $4.6$ for SPM and TLE interactions, respectively. Given the added complexity of implementing the one-way repeater relative to the $CZ$ gate, these small ratios underscore the potential benefit of direct transformations over single- and two-qubit decompositions. Furthermore, this benefit is significantly larger for the implementation with TLEs compared to SPM interactions.

Proposals of spatially distributed architectures relying on nonlinearities from three-level $\Lambda$-systems are found in, e.g., Refs.~\cite{Bartlett2020, basani2025universal}. Their advantage stems from the possibility of using passive nonlinear elements, though circulators are required, and, in some cases, phase-conjugating mirrors to provide time-reversal of wave packets~\cite{basani2025universal}. Importantly, avoiding fidelity degradation due to wave-packet distortion in passive three-level system interactions requires that the pulse bandwidths be significantly smaller than the light-matter coupling rate, $\Gamma_{\rm NL}$. Since spatially distributed nonlinear elements must be traversed sequentially in these architectures, the duration of each layer must be significantly longer than $1/\Gamma_{\rm NL}$. This requirement severely limits the achievable process duration.  
In contrast, the RQPN architecture does not restrict any system process to be slower than $1/\Gamma_{\rm NL}$. 
Furthermore, the effective circuit depth of the RQPN is determined by the duration of the control signals, which can be easily reprogrammed without requiring hardware modifications. Additionally, circulators are straightforwardly avoided by connecting the nonlinear cavities to the input-output channels via, e.g., fast multiport routers~\cite{Heuck2023}.

A proposal for a minimal-resource architecture is found in Ref.~\cite{Bartlett2021}. It utilizes a single nonlinear element and a storage channel to enable counter-propagating wave packets to interact sequentially. With the RQPN architecture, we combine the hardware efficiency of reusing physical elements~\cite{Bartlett2021} with the time efficiency of parallel processing across multiple spatial channels~\cite{steinbrecher2019quantum, ewaniuk2023imperfect, basani2025universal, Bartlett2020}.

\section{Towards Experimental Realizations}\seclab{Towards Experimental Realizations}
In the following, we relate our model assumptions and performance metrics to state-of-the-art demonstrations of integrated photonic circuits and nonlinear systems. 
As a photonic control layer, we consider Pockels materials, such as lithium niobate (LN)~\cite{Zhu2021}, lithium tantalate~\cite{Wang2024}, and barium titanate~\cite{Chelladurai2025} to be the most promising route toward fast and low-loss control over cavity resonances~\cite{Li2020,Larocque2024} and waveguide couplings~\cite{Zhang2019,Xue2022,Herrmann2024,Rasmussen2025}. The Pockels effect is near-instantaneous~\cite{Zhu2021} relative to the 10-100$\,$GHz range control bandwidth offered by scalable CMOS driving electronics~\cite{Daudlin2025}, which is compatible with low-voltage device designs~\cite{Wang2018,Larocque2024}. Dynamic capture and release of optical pulses has already been demonstrated in relatively large ring-resonator structures~\cite{Zhang2019,Li2025}, while Ref.~\cite{Rasmussen2025} reports progress toward analogous control in smaller photonic-crystal cavities, which are particularly attractive because they can support stronger nonlinear interactions when interfaced with quantum emitters.

By neglecting loss and decoherence in the RQPN model, we drastically reduce the computational overhead of optimizing the control fields. This numerical efficiency enabled us to minimize the total process duration, thereby mitigating these detrimental effects without directly including them in the simulations. To relate our results to experimental demonstrations, consider the record-low loss demonstration in a lithium niobate ring resonator with an intrinsic lifetime of $\tau_{\rm{loss}}=24.5\,\mathrm{ns}$ \cite{Zhu24_TFLN_cavity_lifetime} 
For an $N$-photon state stored for a time $t$, the probability of no photon loss is $P_{\rm no-loss}=\mathrm{e}^{-Nt/\tau_{\rm{loss}}}$. Requiring $P_{\rm no-loss}>0.99$, the repeater durations in Table \ref{tab:repeater_results} imply a required nonlinear interaction rate of $\chi_3/(2\pi)=28\,\mathrm{GHz}$ for the SPM architecture with $N=5$ and $g/(2\pi)=69\,\mathrm{GHz}$ for the TLE architecture with $N=6$. For $\chi^{(3)}$-systems, we are not aware of demonstrations showing such large interaction strengths, and the SPM nonlinearity primarily serves as a benchmarking tool for architecture comparisons. For atom-like systems, InAs quantum dots in GaAs photonic crystal nanocavities have demonstrated coupling rates of $g/(2\pi)=40\,\mathrm{GHz}$~\cite{ota2018a}. However, demonstrating these loss levels and interaction strengths in a single device still poses a formidable experimental challenge. 
Heterogeneous integration is a promising approach to combine a controllable integrated circuit with highly nonlinear materials, including 2D materials~\cite{Abajo2025}, carbon nanotubes~\cite{Ovvyan2023}, III-V quantum dots~\cite{Aghaeimeibodi2018}, hot atomic vapors~\cite{Zektzer2024}, and trapped atoms~\cite{Zhou2024}. Even with experimental progress, we expect some level of heralding built into the system control to be necessary for achieving high-fidelity large-scale operations.

We next assess the assumption that the capture and release of traveling pulses (before and after the recirculating evolution, respectively) can be treated as operations that have negligible infidelity and duration compared to the recirculating evolution. 
For nonlinear cavities, capture and release are generally not exactly perfect, since the optimal time-dependent coupling depends on the photon number of the incident or emitted quantum state. Nevertheless, numerical simulations have confirmed that the resulting operation errors are negligible in practice for Gaussian pulses with temporal widths, $T_{\rm pulse}$, much shorter than the timescale of nonlinear interactions, $1/\Gamma_{\rm NL}$~\cite{heuck2020prl,heuck2020pra,krastanov2022dyncav}. When $T_{\rm pulse}\ll 1/\Gamma_{\rm{NL}}$, the capture and release durations are also much shorter than $1/\Gamma_{\rm{NL}}$, which is significantly shorter than the full recirculating evolution of $9$-$18$ units of $1/\Gamma_{\rm NL}$ for our repeater implementations.
Realizing this short-pulse capture and release requires cavity-control bandwidths, $\Omega_{\rm mod}$, to be around $1/T_{\rm pulse}$, i.e., $\Omega_{\rm mod}$ must be larger than $\Gamma_{\rm NL}$, which is straightforwardly realized in Pockels material-based platforms~\cite{Zhu2021}. 

Our RQPN model that describes the recirculating circuit is based on the SLH framework, which assumes a negligible propagation time, $T_{\rm prop}$, of photons traversing the mixing circuit in~\figref{RQPN}(a). For this assumption to be valid, the cavity-waveguide couplings, nonlinear interaction rate, and control modulation bandwidths must be much smaller than $1/T_{\rm prop}$. To estimate $T_{\rm prop}$, we consider a directional coupler length of $L \simeq 100\,\upmu$m and waveguide group index of $n_g\simeq 2.3$. In our repeater example, $T_{\rm prop}=2\!\times\! n_g L/c \simeq 1.5\,\mathrm{ps}$ and $1/T_{\rm prop} \simeq (2\pi)\!\times\!106\,$GHz, while the couplings and control bandwidths were below $8g$ (see Supplemental Material SVI~\cite{grovn2026supplementary}, Fig.~S9). 
Using $g/2\pi=40\,$GHz~\cite{ota2018a}, this comparison highlights a tradeoff: while large nonlinear interaction rates are important for reducing the process duration and thereby keeping loss errors small, they can also bring the system dynamics into a regime where finite propagation delays affect the accuracy of the SLH model. Quantifying the impact of finite propagation times requires an explicit delay-aware model~\cite{bundgaard2025waveguideqed}, which we leave for future work.

\section{Conclusion and Outlook}\seclab{Conclusion and Outlook}
The considerations in~\secref{Towards Experimental Realizations} highlight the need to develop architectures that reduce the experimental requirements for practical deterministic photonic QIP. By introducing the RQPN architecture, we have shown that the direct realization of multi-photon interactions significantly reduces the process duration of key QIP transformations, including a sevenfold speedup relative to state-of-the-art photonic architectures. Our results demonstrate that the dynamic control over caviWty interactions offered by RQPNs provides a versatile and advantageous tool for minimizing the resource overhead required for QIP, both in terms of duration and component count. \\  

To evaluate the performance of specific implementations, future work should verify the performance in the presence of loss, decoherence, and finite propagation delays. We expect built-in heralding schemes to be necessary for near-term experimental demonstrations of high-fidelity QIP tasks. 
Incorporating such schemes into the optimal-control framework for RQPNs provides a promising direction for combining heralded operations based on measurement-induced nonlinearities with deterministic nonlinear interactions subject to loss and decoherence.

\section*{Code availability}
The data and code that support the findings of this article are openly available in Ref.~\cite{grovn2026codeRQPN}.

\begin{acknowledgments}
We thank Stefan Krastanov for many helpful discussions.
This work was supported by the Villum Foundation (QNET-NODES, Grant No. 37417), Innovation Fund Denmark (EQUAL, Grant No. 4356-00007B), and the Danish National Research Foundation (NanoPhoton, Grant No. DNRF147).
\end{acknowledgments}

\bibliography{Bibliography}

\end{document}